\documentclass[aps,prd,twocolumn,showpacs,eqsecnum,nofootinbib]{revtex4}

\usepackage{epsfig}
\usepackage{graphicx}
\usepackage{dcolumn}
\usepackage{amsmath}
\usepackage{enumerate}
\usepackage{epstopdf} 

\usepackage{amssymb}
\newcommand{\mev}{~\mathrm{MeV}}
\newcommand{\gev}{~\mathrm{GeV}}
\newcommand{\tev}{~\mathrm{TeV}}
\newcommand{\trh}{\ensuremath{T_{RH}}}
\begin{document}

\title{ Constraints on Minimal SUSY models with warm dark matter neutralinos}
\author{Graciela Gelmini}
\author{Carlos E. Yaguna}
\affiliation{Department of Physics and Astronomy, UCLA,
 475 Portola Plaza, Los Angeles, CA 90095, USA}


\vspace{6mm}
\renewcommand{\thefootnote}{\arabic{footnote}}
\setcounter{footnote}{0}
\setcounter{section}{1}
\setcounter{equation}{0}
\renewcommand{\theequation}{\arabic{equation}}

\begin{abstract}
 If the energy density of the Universe before nucleosynthesis is  dominated by a scalar field $\phi$ that decays and reheats the plasma to a low reheating temperature $T_{RH}$,  neutralinos may be warm dark matter particles. We study this possibility and derive the conditions on the production mechanism and on the supersymmetric spectrum for which it is viable.  Large values of the $\mu$ parameter   and of the slepton masses are characteristic features of these models.  We compute the expected direct detection cross sections and point out that Split-SUSY  provides a natural framework for neutralino warm dark matter.\end{abstract}


\maketitle
Thermally produced neutralinos are  typical and well-motivated cold dark matter candidates (see e.g.~\cite{Griest:2000kj}). They are produced by scatterings in the thermal bath, then reach  equilibrium and finally decouple when non-relativistic. Non-thermally produced neutralinos, on the contrary, are usually produced as relativistic final states in the decay of heavy particles and  may never reach chemical or kinetic equilibrium. As a result, they are not necessarily cold. Indeed,  if they manage to keep most of their initial energy they might behave as warm dark matter suppressing the evolution of small-scale structures in the Universe.

Although consistent with the observations of the large scale structure of the Universe  and the cosmic microwave background radiation anisotropies, cold dark matter models seem to have problems at galactic scales. They not only tend to form cuspy structures in the halo density profile~\cite{cuspy} but also predict a large overabundance of small halos near galaxies such as our own~\cite{dwarf}. Warm dark matter, with its larger free-streaming scale, may solve these problems while maintaining the celebrated success of cold dark matter models at large scales~\cite{warm}.

 Neutralinos as warm dark matter candidates were initially discussed in Refs.~\cite{Lin:2000qq, Hisano:2000dz}. In this paper we will extend, in several ways, the analysis presented in those references. Ref.~\cite{Lin:2000qq} concentrated on the cosmological aspects of this possibility, while we study a particular particle model. Reheating temperatures smaller than $5\mev$ or $2\mev$ as those found in  Ref.~\cite{Hisano:2000dz} are hardly compatible with the standard cosmological scenario, which require  $T_{RH} > 4\mev$~\cite{Hannestad:2004px}. We, instead take $T_{RH}\simeq10\mev$ as a characteristic value and find the conditions on the initial energy and on the supersymmetric spectrum for which the neutralino is a  warm dark matter particle. In doing so, we  ignore, following Ref.~\cite{Arkani-Hamed:2004fb}, the naturalness argument giving up the idea that SUSY stabilizes the weak scale, notion that is the basis for Split-SUSY models. We also discuss the non-thermal production of neutralinos, as well as  the implications for direct dark matter searches of neutralino warm dark matter.

We concentrate on non-standard cosmological models  (see for example Ref.~\cite{models}) in which the late
 decay of a scalar field $\phi$ reheats the Universe to a low reheating temperature \trh, smaller than the standard neutralino freeze out temperature.  Such scalar fields are  common
in superstring models where they appear as moduli fields. These fields get mass
at the low energy supersymmetry breaking scale, typically of the order of $10^2-10^3$~TeV. The decay of
$\phi$ into radiation increases the entropy, diluting the
neutralino number density. The decay of $\phi$ into supersymmetric
particles, which eventually decay into neutralinos, increases the neutralino
number density. We denote by $b$ the net number of neutralinos
 produced on average per $\phi$ decay.
 The number $b$  is highly
model-dependent, so is the $\phi$ field mass $m_\phi$.  They are determined by the physics of the hidden sector,
 by the mechanism of supersymmetry breaking, and in superstring-inspired
  models by the compactification mechanism \cite{models,gg} . Thus, here  we consider $b$ and $m_\phi$ as  free parameters.

 To account for the dark matter of the Universe, the neutralino relic density must be in agreement with the observed dark matter density. 
   In low $T_{RH}$ cosmological models essentially  all 
 neutralinos   can have the dark matter density
 provided the 
  right combination of the following  two parameters can be achieved in the high energy theory:
 the reheating temperature \trh, and the ratio of the
 number of neutralinos produced per $\phi$ decay over the $\phi$ field mass, i.e.
$b/m_\phi$~\cite{gg, ggsy}.  We will find later the values of \trh~ and $b/m_\phi$ for which the neutralinos we are interested in have the right dark matter density.
 
  A crucial quantity that distinguishes warm from cold dark matter is the free-streaming length at matter-radiation equality $\lambda_{FS}$,
which depends on the parameter $r_\chi=a(t)p_\chi(t)/m_\chi$ (see for example Refs.~\cite{Lin:2000qq, Hisano:2000dz} and references therein). This parameter would be the present characterictic speed of  neutralinos of mass $m_\chi$, if their momentum $p_\chi$ only redshifted from neutralino production onwards  ($a(t)$ is the scale factor with $a_0 =1$). During the cosmic evolution $r_\chi$ is constant. Structures smaller than $\lambda_{FS}$ are damped because neutralinos can freely flow out of them. N-body simulations have shown that to explain the lack of substructure in the local group, $\lambda_{FS}$ should be of order $0.1~\mathrm{Mpc}$~\cite{warm}, thus $r_\chi\simeq10^{-7}$.

The  parameter  $r_\chi$ also determines (for neutralinos which are relativistic at production~\cite{Cembranos:2005us}) the neutralino phase-space density $Q$. This is
defined  as $Q= \rho/ \langle v^2 \rangle^{3/2}$ where $\rho$ is the neutralino energy density and
$\langle v^2\rangle$ is the mean square value of the particle velocity. In the absence of dissipation,
the coarse-grained phase-space density (the quantity that can actually be observed)
can only decrease from its primordial value. The observation of dwarf-spheroidal galaxies places a lower bound on $Q$ which translates into an upper bound on
$r_\chi$ of about 2.5$\times 10^{-7}$~\cite{Lin:2000qq}. In what follows we impose
 the condition $r_\chi =10^{-7}$ to our model.

Neutralinos are produced in $\phi$ decays, thus  their average initial energy is 
 $E_I=m_\phi/N$, where $N$ is a number which depends on the production spectrum. We expect  $N$ to be  of order one and require  $E_I \gg m_\phi$ so that neutralinos are relativistic at production. Thus,  $p_\chi \simeq E_I$ at the moment of $\phi$ decay.
 Assuming an instantaneous  $\phi$ decay at   \trh, with no subsequent entropy production, the scale factor at decay is $a = T_0/ T_{RH}$, where  $T_0$ is the present photon temperature, and the parameter $r_\chi$  in our model  is $r_\chi=(T_0 E_I)/ (T_{RH} m_\chi)$ i.e.
\begin{equation}
r_\chi \simeq  10^{-7} \left(\frac {2.3}{N}\right) \left(\frac{m_\phi}{10^3 \tev}\right) \left(\frac{10\mev}{\trh}\right) \left(\frac{100~\mathrm{GeV}}{m_\chi}\right).
\label{eq:rchi}
\end{equation}
Thus,  the condition
 $r_\chi =10^{-7}$ fixes $m_\phi$ in terms of the reheating temperature and the neutralino mass
\begin{equation}
m_\phi= 10^3{\rm TeV} \left(\frac {N}{2.3}\right)\left(\frac{m_\chi}{100{\rm GeV}}\right)\left(\frac{T_{RH}}{10{\rm MeV}}\right).
\label{eq:mphi}
\end{equation}

In the estimation of  $r_\chi$ we  have assumed that
the neutralinos do not lose their energy in scattering processes with the thermal bath. To ensure this condition we will simply require that the interactions of neutralinos with the particles present in the plasma, $e^\pm$,$\nu$ and $\gamma$, are out of equilibrium. Neutralino interactions are determined by the neutralino  composition in terms of  gauge eigenstates.  The ligthest neutralino ($\chi$) is a linear superposition of bino ($\tilde B^0$), wino ($\tilde W^0$) and higgsino ($\tilde H^0$) states,
\begin{equation}
\chi=N_{11}\tilde B^0+N_{12}\tilde W^0+N_{13}\tilde H_d^0+N_{14}\tilde H_u^0\,. 
\label{eq:chi}
\end{equation}
 According to the dominant term in Eq.~(\ref{eq:chi}), $\chi$ is classified as bino-like, wino-like, or higgsino-like. 
 
 Wino and higgsino-like neutralinos are always accompanied by a chargino state ($\chi^{\pm}$) with a mass difference $\Delta m= m_{\chi^{\pm}}- m_\chi \simeq \sin^2(\theta_W) (M_Z/m_\chi)^2 m_\chi$ for higgsinos and much smaller for winos.
 When $m_\chi \Delta m < E_I T$ relativistic  neutralinos   scatter inelastically (e.g. $\chi e^-\rightarrow \chi^- \nu$)  besides  scattering elastically (e.g. $\chi \nu\rightarrow \chi \nu$) on electrons and neutrinos.  Since it is not possible to simultaneously suppress the elastic and the inelastic scattering rates, wino and higgsino-like neutralinos easily lose their initial energy.  In Ref.~\cite{Hisano:2000dz}   was indeed pointed-out that a wino-like neutralino cannot be warm dark matter and that a higgsino-like neutralino would only be viable if $T_{RH}< 2\mev$ were allowed.  Bino-like neutralinos, on the other hand, couple to electrons and neutrinos mainly through slepton exchange and typically have  small interaction cross sections. In general, there are no charginos close in mass to bino-like neutralinos and inelastic scatterings can be suppressed. We will therefore focus on bino-like neutralinos in the following.

In the approximation $m_{\tilde \ell} \gtrsim  6T E_I$, which is guaranteed by the lower bound found below, the cross section for elastic bino-lepton ($\chi$-$\ell$) scattering mediated by a slepton of mass $m_{\tilde \ell}$ is estimated to be
\begin{equation}
\sigma v\simeq \frac{g^4}{16\pi}~ \frac{E_{\ell i} E_{\ell f}}{m_{\tilde \ell}^4}.
\label{eq:sigv}
\end{equation}
 Here $g$ is the weak hypercharge coupling constant, $E_{\ell i}\simeq 3 T$ is the energy of the incoming thermal-bath neutrino or electron, and  $E_{\ell f}$ is the energy of the outgoing lepton. We require that $\chi$-$\ell$ interactions are out-of-equilibrium, i.e. that the interaction rate is smaller than the expansion rate of the Universe, $\Gamma \simeq n\sigma v< H$, where $n$ is the thermal neutrino or electron number density and $H$ is the Hubble parameter. Since the interaction rate goes as $T^5$ and $H$ as $T^2$, if  the interactions are out-of-equilibrium at $T_{RH}$  they will also be out of equilibrium at any later temperature $T<T_{RH}$. Replacing 
 $E_{\ell f}$ in Eq.~\ref{eq:sigv} by its maximum value $E_I=m_\phi/N$, we obtain that neutralino interactions are out of equilibrium at $T_{RH}$ (and subsequently) if
\begin{equation}
0.1 \left(\frac{m_\phi}{N 10^3\mathrm{TeV}}\right) \left(\frac{T_{RH}}{10~\mathrm{MeV}}\right)^2\left(\frac{40~\mathrm{TeV}}{m_{\tilde\ell}}\right)^4 < 1.
\label{eq:snu}
\end{equation}
Sleptons therefore must be  heavy.
 
Due to the structure of the neutralino mass matrix, bino-like neutralinos always have a small higgsino component. As a result, they also couple to fermions through $Z$ exchange. If sleptons are  heavy, the $Z$-exchange  diagram could give the dominant contribution to the scattering rate. In this case the neutralino cross section 
 is
\begin{equation}
\sigma v\lesssim \frac{g^4}{32\pi}~ \tan^4\theta_W\cos^22\beta~ \frac{E_{\ell i} E_{\ell f} }{\mu^4}\,,
\label{eq:sigma2}
\end{equation} 
where $\theta_W$ is the weak mixing angle and $\tan\beta=v_2/v_1$ is the ratio between the vacuum expectation values of the two Higgs doublets.
Eq (\ref{eq:sigma2}) holds for   $m_Z^2  \lesssim 6T E_I$  and    $m_\chi^2 \lesssim 6T E_I$.
 The latter condition,
 after using Eq.~\ref{eq:mphi}, becomes
$T_{RH} /10 MeV \gtrsim  1.3 ({m_\chi}/{100 {\rm GeV}})$.
The condition  $\Gamma <H$ implies for the interaction in Eq.~\ref{eq:sigma2} (again replacing $E_{\ell f}$ by its maximum possible value, i.e. $E_I$)
\begin{equation}
0.1\left(\frac{m_\phi}{N10^3\tev}\right)\left(\frac{T_{RH}}{10\mev}\right)^2\left(\frac{8~\mathrm{TeV}}{\mu}\right)^4<1\,.
\label{eq:mu}
\end{equation}
If $m_\chi^2 \gtrsim 6T E_I$ (i.e.
$T_{RH} /10 MeV \lesssim  1.3 ({m_\chi}/{100 {\rm GeV}})$, the right-hand side of Eq.~\ref{eq:sigma2} and the left-hand side of Eq.~\ref{eq:mu} must be multiplied by the factor $m_\chi/ 3T_{RH} E_I$, which yields
\begin{equation}
0.1\left(\frac{m_\chi}{100{\rm GeV}}\right)^2\left(\frac{T_{RH}}{10\mev}\right)\left(\frac{6~\mathrm{TeV}}{\mu}\right)^4<1\,.
\label{eq:mu2}
\end{equation}
In any event, a large  $\mu$ parameter value is required.

Replacing  Eq.~\ref{eq:mphi} into Eqs.~\ref{eq:snu} and \ref{eq:mu}
and directly from Eq.~\ref{eq:mu2} we obtain bounds of the slepton masses and the parameter $\mu$ which depend only on the neutralino mass and the reheating temperature,
\begin{equation}
{m_{\tilde\ell}} > 18 {\rm TeV}  \left(\frac{m_\chi}{100{\rm GeV}}\right)^{1/4}\left(\frac{T_{RH}}{10{\rm MeV}}\right)^{3/4},
\label{eq:snu2}
\end{equation}
\begin{equation}
\mu > 4 {\rm TeV}  \left(\frac{m_\chi}{100{\rm GeV}}\right)^{1/4}\left(\frac{T_{RH}}{10{\rm MeV}}\right)^{3/4},
\label{eq:mu3}
\end{equation}
from Eq.~\ref{eq:mu} or
\begin{equation}
\mu > 3 {\rm TeV}  \left(\frac{m_\chi}{100{\rm GeV}}\right)^{1/2}\left(\frac{T_{RH}}{10{\rm MeV}}\right)^{1/4}.
\label{eq:mu4}
\end{equation}
when Eq.~\ref{eq:mu2} holds instead.

Using the cosmological lower bound $T_{RH} > 4$ MeV we obtain ${m_{\tilde\ell}} > 9~{\rm TeV}$  $(m_\chi/100 {\rm GeV})^{1/4}$ and $\mu $ $> 2.0~{\rm TeV}$  $(m_\chi/100 {\rm GeV})^{1/4}$ or
$\mu $ $> 2 ~{\rm TeV}$  $(m_\chi/100 {\rm GeV})^{1/2}$.

 Let us now return to the density requirement. In general there are four different ways in which the neutralino  relic density
$\Omega h^2$ depends on $\trh$, in the low reheating cosmologies we are considering. The four cases have  thermal or non-thermal
dominant neutralino production, with or without chemical equilibrium~\cite{gg,  ggsy}. The case of interest here is that of a very low reheating temperature and  high standard neutralino relic density (corresponding to  small neutralino cross sections) for which the production is non-thermal without chemical equilibrium~\cite{gg}. In this case neutralinos are produced
in the decay of the $\phi$ field into supersymmetric particles, with subsequent fast decay of all other supersymmetric particles into neutralinos, and the production is not compensated by annihilation.
This implies
\begin{equation}
\frac{\Omega_\chi h^2}{0.11}\simeq 2\,10^3 b \left(\frac{10^3\mathrm{TeV}}{m_\phi}\right) \left(\frac{m_\chi}{10^2{\rm GeV}}\right)\left(\frac{T_{RH}}{10 \mathrm{MeV}}\right).
\label{eq:omega}
\end{equation}
 This equation fixes the value of $b/ m_\phi$
required for neutralinos to have the dark matter density, 
$\Omega_{\rm dm} h^2=0.11$~\footnote{$\Omega_{\rm dm} h^2=$ 
$0.109^{+0.003}_{-0.006}$ was obtained for a $\Lambda$CDM model with scale-invariant primordial perturbation spectrum through a global fit of cosmic microwave background, supernovae, and large scale structure data~\cite{wmap2}},
\begin{equation} 
\frac{b}{m_\phi} \simeq \left(\frac{5\,10^{-7}}{{\rm TeV}}\right) \left(\frac{m_\chi}{100 {\rm GeV}}\right) \left(\frac{10{\rm MeV}}{T_{RH}}\right).
\label{eq:b}
\end{equation}

\begin{figure}
\includegraphics[scale=0.3]{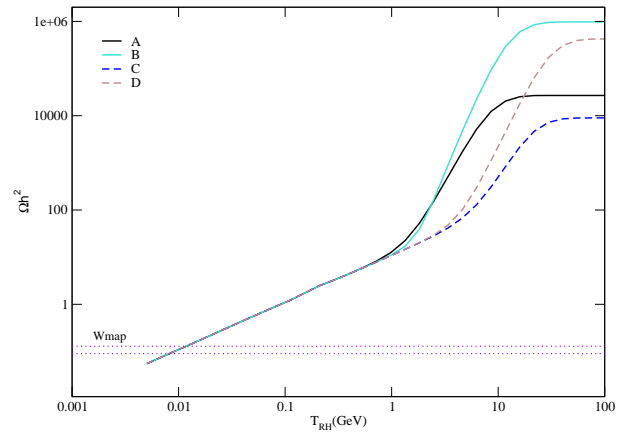}
\caption{The neutralino relic density as a function of $T_{RH}$ in four different models compatible with warm dark matter. In all of them $M_2,M_3,m_A=1\tev$, $\tan\beta=10$, $A_t,A_b=0$. In A, $m_\chi \simeq M_1=100\gev$, $\mu=10 \tev$, $m_{\tilde f}=50\tev$, and $b/m_\phi=4\times 10^{-7}\tev^{-1}$.  In B, $m_\chi \simeq M_1=100\gev$, $\mu,m_{\tilde f}=100\tev$, and $b/m_\phi=4\times 10^{-7}\tev^{-1}$.  In C, $m_\chi \simeq M_1=300\gev$, $\mu=10\tev$, $m_{\tilde f}=50\tev$, and $b/m_\phi=1.3\times 10^{-7}\tev^{-1}$. In D, $m_\chi \simeq M_1=300\gev$, $\mu, m_{\tilde f}=100\tev$, and $b/m_\phi=1.3\times 10^{-7}\tev^{-1}$.}
\label{fig:trh}
\end{figure}

Using the program described in Ref.~\cite{ggsy} we have followed the evolution of the relic density of some neutralinos  warm dark matter  as function of \trh. This is shown in 
Fig.~\ref{fig:trh}. We use a Minimal Supersymmetric Standard Model defined in terms of the parameter set $M_{3}$, $M_2$, $M_1$, $m_A$, $\mu$, $\tan\beta$, $m_0$, $A_t$, and $A_b$. Here $M_{i}$ are the three gaugino masses, $m_A$ is the mass of the pseudoscalar higgs boson, and $\tan\beta$ denotes the ratio $v_2/v_1$. The soft breaking scalar masses are defined through the simplifying ansatz $M_Q=M_U=M_D=M_E=M_L=m_0$ whereas the trilinear couplings are given by $A_U=\mathrm{diag(0,0,A_t)}$, $A_D=\mathrm{diag(0,0,A_b)}$, and $A_E=0$. All these parameters are defined at the weak scale. The models presented in Fig.~\ref{fig:trh} have $m_\chi \simeq M_1=100$ GeV (solid lines) or 300 GeV (dashed lines) and other parameters as mentioned in the figure caption. 

Fig.~\ref{fig:trh} 
confirms that at low reheating temperatures the neutralino production is purely non-thermal,  independently of  the supersymmetric spectrum, and follows Eq.~(\ref{eq:omega}).  Thermal effects are relevant only for $T_{RH}\gtrsim 1\gev$. The neutralino relic density shown in the figure correspond to  $b/m_\phi=4\times 10^{-7}\tev^{-1}$ for $m_\chi=M_1=100\gev$ (solid lines),  and to $b/m_\phi=1.3\times 10^{-7}\tev^{-1}$ for $m_\chi=M_1=300\gev$.
  
Combining Eqs.~\ref{eq:rchi} and  \ref{eq:omega},  the requirement of neutralinos to account for the whole of the dark matter and to be warm dark matter, fixes $b$ and $m_\phi$ (see Eq.~\ref{eq:mphi}) separately,
\begin{equation}
\frac{\Omega_\chi h^2}{0.11}\simeq 5\,10^3  \left(\frac{b}{N}\right) \left(\frac{10^{-7}}{r_{\chi}}\right).
\label{eq:omega2}
\end{equation}
That is, fixing $\Omega_\chi h^2=0.11$ and  $r_\chi=10^{-7}$ requires $b= 2~ 10^{-4}~N$. 

Neutralinos, therefore, can be warm dark matter particles only if slepton masses and $\mu$ are both  large, above about 10 and 2 TeV respectively. Supersymmetric models with heavy sfermions do exist in the literature. In Split-SUSY models~\cite{Arkani-Hamed:2004fb,Arkani-Hamed:2004yi, Giudice:2004tc}, all scalar superpartners are usually heavy, with a mass scale that, in principle, could go up to the GUT or the Planck scale. These models solve a number of phenomenological problems associated with ordinary supersymmetry: they alleviate proton decay, increase the mass of the Higgs boson, and minimize the flavor and CP problems. Thus, a Split-SUSY model with a bino-like LSP and $\mu\gtrsim 10\tev$ would be a perfectly viable framework for neutralino warm dark matter.

   In standard cosmological models, a bino-like neutralino is allowed in Split-SUSY models only if it is almost degenerate in mass with other neutralino and chargino states ($M_1\simeq M_2$ or $M_1\simeq \mu$) and coannihilation effects determine its relic density~\cite{Masiero:2004ft}. These constraints, however, do not hold in our model. By assuming a non-standard cosmology which allows for non-thermal production of neutralinos we have effectively enlarged the viable parameter space of Split-SUSY models.

In Split-SUSY models all scalar superpartners are usually heavy but our conditions require only that the sleptons be heavy, not necessarily the squarks. A model with light squarks would certainly be appealing from a phenomenological point of view. The most general low energy spectrum of masses below a TeV  of a model compatible with neutralino warm dark matter would consist of: the lightest neutralino, the next-to-lightest neutralino, the ligthest chargino, the three Higgs bosons,  the gluino and the squarks. 
Such low energy spectrum is sufficiently rich to give significant signals at the LHC, particularly for light gluinos and squarks. Besides, the absence of sleptons in the low energy spectrum would be a crucial test for models with neutralino warm dark matter.  
\begin{figure}
\includegraphics[scale=0.3]{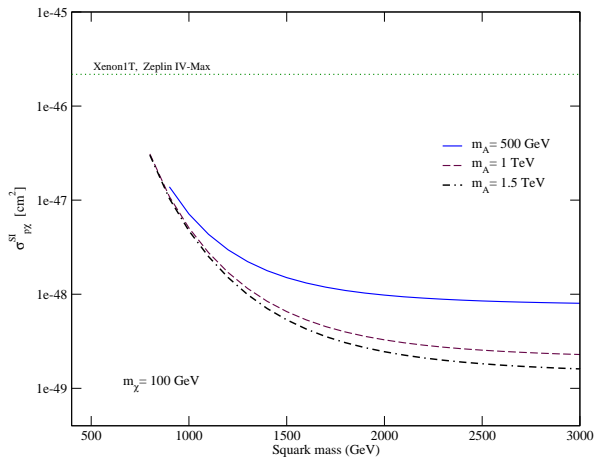}
\caption{Spin-independent neutralino-proton cross section for warm dark mater neutralinos as function of the squark mass, for different values of $m_A$, and for $M_1=m_\chi=100$GeV,  $M_2=M_3= $1TeV, $\mu=10$TeV, slepton masses of 50TeV and $\tan\beta=10$.   The horizontal dotted line shows the reach of Zeplin IV-MAX and Xenon 1T.}
\label{fig:detection}
\end{figure}

The prospects for  direct detection of bino-like neutralinos depend on
the neutralino-nucleon cross section, which in turn is determined by the
diagrams mediated by the heavy neutral  Higgs boson (H) and  squarks. We
show in Fig.~\ref{fig:detection} the spin-independent neutralino-proton
scattering cross section as a function of the squark mass for  three
different models compatible with neutralino warm dark matter, each with a different value of $m_A$, the pseudoscalar Higgs boson mass. The mass of  heavy boson H is close to $m_A$.
The lines are not continued to smaller squark masses because they
enter into a region
incompatible with the higgs mass bound \cite{unknown:2006cr} or $b\rightarrow s\gamma$ \cite{Koppenburg:2004fz}.  The figure shows
that the expected neutralino-nucleon scattering cross sections in neutralino warm dark matter models are small, between
$10^{-46}~ cm^{2}$ and $10^{-49}~ cm^{2}$.  Hence, the expected
spin-independent neutralino-proton cross sections are below the reach  of
the largest dark matter detectors envisioned at present, i.e. Zeplin IV-MAX
and Xenon 1 Ton~\cite{Gaitskell:2004gd}.

In this letter we studied  non-thermally produced neutralinos that are
viable warm dark matter candidates, in low reheating temperature cosmological scenarios. We have shown that models compatible with neutralino warm
dark matter satisfy a number of non-trivial requirements, some of them testable:
the reheating temperature is low, the mass of the decaying scalar field which reheats the Universe  and the number of  neutralinos
it produces per decay are constrained, the neutralino is bino-like,  slepton masses
and the mu parameter are both large  and the neutralino interacts too
weakly to be observed in  direct dark matter experiments in the near future.  To validate the idea of
neutralino warm dark matter, therefore,  cosmological observations as well
as accelerator and dark matter searches will be essential.

This work was supported in part by the US Department of Energy Grant
DE-FG03-91ER40662, Task C  and NASA grants NAG5-13399  and ATP03-0000-0057.
We thank P. Gondolo for helpful suggestions.

\end{document}